\documentclass[12pt,reqno]{amsart}

\newtheorem{theorem}{Theorem}
\newtheorem{lemma}[theorem]{Lemma}

\newtheorem{definition}[theorem]{Definition}

\theoremstyle{remark}

\def\RE{\mathbb R}
\def\LD{L^2(\RE^3)}

\def\HUB{\bar H^{1,2}(\RE^3)}
\def\HDB{\bar H^{2,2}(\RE^3)}
\def\C{{\mathbb C}}
\def\supp{\text{\rm supp\,}}
\def\diam{\text{\rm diam}}
\def\dist{\text{\rm dist}}
\def\G{\mathcal G}

\begin{document}

\title[Finite speed of propagation and local boundary conditions]
{Finite speed of propagation and local boundary conditions for
wave equations with point interactions}

\author{Pavel Kurasov}

\address{Department of Mathematics, Lund Institute of Technology,
P.O. Box 118, 22100 Lund, Sweden}

\email{kurasov@maths.lth.se}

\author{Andrea Posilicano}

\address{Dipartimento di Scienze, Universit\`a dell'Insubria, I-22100
Como, Italy}

\email{posilicano@uninsubria.it}

\begin{abstract} We show that the boundary conditions entering in the
definition of the self-adjoint operator $\Delta^{A,B}$ describing
the Laplacian plus a finite number of point interactions are local
if and only if the corresponding wave equation
$\ddot\phi=\Delta^{A,B}\phi$ has finite speed of propagation.
\end{abstract}

\maketitle

\section{Introduction and framework}

The Laplace operator with point interactions in $ L^2(\RE^3) $ can
be defined in the following way (see \cite{[AGHH],[AK]} and
references therein for more details). Consider any finite set $ Y
= \{ y_j \}_{j=1}^n $ of points from $ \RE^3.$ Then the
restriction $ - \Delta_0 $ of the Laplace operator $ - \Delta $ to
the set of functions from the Sobolev space $ H^{2,2} (\RE^3) $
vanishing at all points $ y_j\in Y $ is a symmetric operator with
deficiency indices $ (n,n).$ The domain of the adjoint operator $
- \Delta_0^* $ coincides with $ H^{2,2} (\RE^3 \setminus Y)$.
Every function $ \phi $ from $ H^{2,2} (\RE^3\setminus Y) $
possesses the following asymptotic representation in a
neighborhood of each point $ y_j \in Y $
\begin{equation}
\phi (x) = \frac{1}{4 \pi \vert x-y_j\vert} \phi_j^{s} +
\phi_j^{r} + O(\vert x-y_j\vert), \; \; x \rightarrow y_j,
\end{equation}
where the coefficients $ \phi_j^{r} $ and $ \phi_j^{s} $ can be
considered as certain generalized boundary values of the function
$ \phi .$ These boundary values can be used to describe all
self-adjoint extensions of $ -\Delta_0 $ as restrictions of the
adjoint operator $ -\Delta_0^* $ to the set of functions
satisfying the generalized boundary conditions:
\begin{equation} \label{bc}
A \vec{\phi}^{r} = B \vec{\phi}^{s}.
\end{equation}
The $ n$-dimensional complex vectors $ \vec{\phi}^{s} $ and $
\vec{\phi}^{r} $ have coordinates $ \phi^{s}_j $ and $ \phi^{r}_j
$ respectively and the $ n \times n $ matrices $ A , B $ satisfy
\begin{equation} \label{symm}
AB^* = BA^*,
\end{equation}
and
\begin{equation} \label{rang}
{\rm rank}\,(A, B)= n\,, \end{equation} where $ (A, B) $ is the
$n\times 2n$ matrix obtained from $ A $ and $ B.$ The boundary
form for the operator $ -\Delta_0^* $
$$ \langle (-\Delta_0^*) \phi,\psi \rangle_{L^2} -
\langle \phi, (-\Delta_0^*) \psi \rangle_{L^2} $$ determines the
following symplectic form in the space $ \mathbb C^{2n} \ni
\vec{\phi} \equiv (\vec{\phi}^s, \vec{\phi}^r) $ of boundary
values:
\begin{equation}
\left[ \vec{\phi}, \vec{\psi} \right]_n := \left\langle \left(
\begin{array}{c}
\vec{\phi}^r \\
\vec{\phi}^s \end{array} \right), \left( \begin{array}{cc}
0 & I_n \\
-I_n & 0 \end{array} \right) \left( \begin{array}{c}
\vec{\psi}^r \\
\vec{\psi}^s \end{array} \right) \right\rangle_{\C^{2n}}.
\end{equation}
Then all self-adjoint extensions of $ -\Delta_0 $ can be described
by Lagrangian planes  associated with this symplectic form. Every
such plane $ \mathcal L$ is described by (\ref{bc}) provided that
the matrices $ A,B $ satisfy conditions (\ref{symm}) and
(\ref{rang}). The first condition  guarantees that the symplectic
form vanishes for all $ \vec{\phi}, \vec{\psi} \in \mathcal L ,$
i.e. that the corresponding extension of $ -\Delta_0 $ is
symmetric. The second condition guarantees that the plane $
\mathcal L $ has the maximal dimension $ n $, i.e. that the
operator extension is not only symmetric but self-adjoint as well.

\begin{definition} \label{def1} The operator
$ - \Delta^{A,B} $ is the restriction of the adjoint operator $ -
\Delta_0^* $ to the domain
\begin{equation}
D\, (-\Delta^{A,B}) = \left\{ \phi \in H^{2,2} (\RE^3 \setminus Y
)\,:\, A \vec{\phi}^{r} = B \vec{\phi}^{s} \right\}.
\end{equation}
\end{definition}

It is clear that different pairs of matrices $ (A, B) $ can
determine the same self-adjoint extension of $ - \Delta_0.$ If the
matrix $ A $ is invertible then the boundary conditions (\ref{bc})
can be written as
\begin{equation} \label{bc2}
\vec{\phi}^{r} = H \vec{\phi}^{s},
\end{equation}
with a Hermitian matrix $ H = A^{-1} B.$ The operator $ - \Delta^H
$ will be identified with the operator $ -\Delta^{I_n,H}$.

The action of the operator $ -\Delta^{A,B} $ coincides with the
action of the (differential)
 Laplace operator and therefore this operator is always local, i.e. $\supp - \Delta^{A,B}
\phi\subseteq \supp\phi$ for any $\phi \in D\,(-\Delta^{A,B})$ and
for any admissible pair $ A,B$. However the boundary conditions
(\ref{bc}) entering in Definition \ref{def1} are local, i.e. do
not connect the "boundary values" $ \phi_j^{s}, \phi_j^{r} $ at
different points $ y_j $ if and only if the matrices $ A $ and $B
$ can be chosen diagonal. In this case to check whether the
function $ \phi $ satisfies the boundary conditions (\ref{bc2}) or
not, one needs to check the behavior of the function $ \phi $ in a
certain small neighborhood of each point $ y_j $ separately.  In
this case the $n$-dimensional Lagrangian plane $ \mathcal L $
corresponding to (\ref{bc}) can be presented as a tensor product
of $n$ $1$-dimensional Lagrangian planes for the symplectic
 form in $\C^2$
\begin{equation}
\left[ \left( \begin{array}{c} {\phi}_j^s\\
 {\phi}_j^r \end{array} \right), \left( \begin{array}{c} {\psi}_j^s\\
  {\psi}_j^r \end{array} \right) \right]_1 :=
\left\langle \left( \begin{array}{c}
{\phi}_j^r \\
{\phi}_j^s \end{array} \right), \left( \begin{array}{cc}
0 & 1 \\
-1 & 0 \end{array} \right) \left( \begin{array}{c}
{\psi}_j^r \\
{\psi}_j^s \end{array} \right) \right\rangle_{\C^2}
\end{equation}
defined for the boundary values $ (\phi^r_j, \phi^s_j) $
associated with the point $ y_j .$ Such boundary conditions and
corresponding self-adjoint extensions of $ - \Delta_0 $ will be
called {\it local}.

In fact all interesting local extensions corresponds to boundary
conditions that can be written in the form (\ref{bc2}) with a
diagonal matrix $ H $. The boundary conditions are local if the
matrices $ A,B $ can be chosen diagonal. These conditions can be
written as (\ref{bc2}) if the matrix $ A $ is invertible which can
be assumed without loss of generality. Indeed if it is not the
case then one can remove few points from the set $ Y $ to get
boundary conditions equivalent to (\ref{bc}) with invertible
matrix $ A$ (may be having a certain smaller dimension).

The purpose of this note is to show (see Theorem \ref{Th3} below)
that the boundary conditions (\ref{bc}) are local if and only if
the wave equation
\begin{equation} \label{we}
\ddot\phi-\Delta^{A,B} \,\phi = 0
\end{equation}
has finite speed of propagation. We remind that in general an
abstract wave equations
\begin{equation} \label{we2}
\ddot\phi + \mathcal A\,\phi = 0
\end{equation}
is said to have finite speed of propagation if for any solution
$\phi$ and for any real $t$,
\begin{equation}\label{lsp}
\diam(\supp \phi(t))\le 2v\,|t|+\diam \left(\supp
\phi(0)\cup\supp\dot\phi(0)\right)
\end{equation}
holds with a certain $v\in (0,\infty)$. While it is obvious what
``local boundary conditions'' means in the case of $ -
\Delta^{A,B}$, the situation is not so clear for  self-adjoint
operators arising from extensions of symmetric operators obtained
by restricting the Laplacian to the  set of smooth functions with
compact support on $\RE^3\setminus M$, where $M$ is not discrete,
for example a subset with Hausdorff dimension $0<d<2$ or a low
dimensional manifold \cite{[AK],pavlov}. Since the wave equation
(\ref{we2}) is always well defined (in the sense that it generates
a strongly continuous group of evolution) for any bounded from
below self-adjoint operator $\mathcal A$ (see e.g \cite{[G]},
chapter 2, section 7), our result could be used to shed light on
the problem of the locality of boundary conditions in more
complicated situations.

 We conclude by pointing
out that locality of boundary conditions, hence finite speed of
propagation, is equivalent to locality in the sense of forms.
According to \cite{[S]}, section 1.2, the self-adjoint operator
$-\Delta^{A,B}$ is said to be form-local if the sesquilinear form
of the operator $ -\Delta^{A,B} $ vanishes on any two functions
from the domain $D(F^{A,B})$ of the quadratic form  having
disjoint supports. The Theorem 2 from \cite{[P]} implies that the
operator $ -\Delta^{A,B} $ is form local if and only if the
boundary conditions (\ref{bc}) can be written in the form
(\ref{bc2}) with local (i.e. diagonal) operator $ H $. Hence the
boundary conditions (\ref{bc2}) with diagonal matrices $ H $
describe all form-local operators.

\section{Proofs}
The following representation for functions from the domain of the
operator $ - \Delta^{A,B}$ will be used in our proofs.

We denote by $\HUB$ the homogeneous Sobolev space of tempered
distributions $\phi$ such that $ i \nabla\phi$ is square
integrable. Then $\HDB$ denotes the space of $\phi\in\HUB$ such
that $ -\Delta\phi\in\LD$. In general neither $\phi\in\HUB$ nor
$\phi\in\HDB$ imply $\phi\in \LD$; by Sobolev embedding theorems
one has $\HUB\subset L^6(\RE^3)$ and $\HDB\subset C_b(\RE^3)$.\par
We use the symbol $\G$ to denote the Green function of $-\Delta$,
i.e.
$$
\G(x):=\frac{1}{4\pi}\,\frac{1}{|x|}\,.
$$
For any  $y_j\in Y$ we define the functions
$$
\G_j(x):=\G(x-y_j)\,,\qquad d_j(x):=|x-y_j|\,,
$$
and the symmetric matrices $G=(G_{ij})$ and $D=(d_{ij})$, $1\le
i,j\le n$, by
$$
G_{ij}:=\G(y_i-y_j)\,,\ i\not=j\,,\ G_{jj}:=0\,, \qquad
d_{ij}:=|y_i-y_j|\,.
$$
\begin{lemma}\label{lemma} The self-adjoint operator $-\Delta^{A,B}$ given in
Definition 1 can be re-written in the following way
\begin{align*}
D( - \Delta^{A,B}):= &
\left\{\,\phi\in\LD\,:\,\phi=\phi_0+\sum_{1\le j\le
n}\zeta_\phi^j\, \G_j, \
\phi_0\in\HDB,\right. \\
 & \left. \quad \vec\zeta_\phi\in\C^n, \sum_{1\leq j\leq n}
A_{ij} \phi_0(y_j)=\sum_{1\le j\le n}
(B-AG)_{ij}\,\zeta_\phi^{j}\, \right\}\,,
\end{align*}
$$
 -\Delta^{A,B}\phi := - \Delta\phi_0\,.
$$
\end{lemma}
\begin{proof} It is sufficient to note that for any
$\phi$ from the operator domain defined above one has $\phi\in
H^{2,2}(\RE^3\setminus Y)$ and $$ \lim_{x\to
y_j}\,(\phi-\zeta^j_\phi\,\G_j)(x)= \phi_0(y_j)+\sum_{1\le k\le
n}G_{jk}\,\zeta_\phi^{k}\,.
$$
Thus
$$ \phi_j^{s}=\zeta^j_\phi\,,\qquad
\phi_j^{r} =\phi_0 (y_j)+ \sum_{1\le k\le
n}G_{jk}\,\zeta_\phi^{k}\,.
$$
\end{proof}
The proof of our main result relies on the following
representation of the solutions of the Cauchy problem given by the
wave equation (2). This result, in the simpler case of a single
point interaction, was already obtained, by a different, less
explicit proof, in \cite{[NP1]} (also see \cite{[NP2]}).
\begin{theorem} \label{Th2}
The Cauchy problem
\begin{equation} \label{problem}
\begin{array}{c} \displaystyle \ddot\phi=\Delta^{A,B}\phi \\
\phi(0)\in D(\Delta^{A,B}) \\
\dot\phi(0)\in D(F^{A,B})
\end{array}
\end{equation}
has an unique strong solution
$$\phi\in C(\RE;D(\Delta^{A,B})\cap
C^1(\RE;D(F^{A,B}))\cap C^2(\RE;\LD)
$$
explicitly given, when $t\ge 0$, by
$$
\phi(t)=\phi_{f}(t)+\sum_{1\le j\le
n}\theta(t-d_j)\,\zeta_\phi^j(t-d_j)\,\G_j\ ,
$$
where $\theta$ denotes the Heaviside function, $\phi_{f}$ is the
unique solution of the Cauchy problem for the free wave equation
\begin{align} \label{solf1}
&\ddot\phi_f =\Delta\phi_f\\
\label{solf2}
&\phi_f(0)=\phi(0)\\
\label{solf3} &\dot\phi_f(0) =\dot\phi(0)\,,
\end{align}
and $\vec\zeta_\phi(t)$, $t\ge 0$, is the unique solution of the
Cauchy problem for the system of inhomogeneous retarded
first-order differential equations
\begin{align} \label{retard}
&\sum_{1\le j\le
n}A_{ij}\left(\frac{\dot\zeta_\phi^j}{4\pi}-\phi_f(y_j)\right) +
B_{ij}\zeta_\phi^{j} = \sum_{1\le j,k\le
n}A_{ij}G_{jk}\theta(\cdot-d_{jk})
\, \zeta_\phi^k(\cdot-d_{jk})\,,\\
\label{retard1} &\zeta_\phi(0)=\zeta_{\phi(0)}\,.
\end{align}
\end{theorem}
\begin{proof} Being $-\Delta^{A,B}$ a bounded from below self-adjoint
operator, it is well known from the theory of abstract wave
equations in Hilbert spaces (see e.g \cite{[G]}, chapter 2,
section 7) that the corresponding Cauchy problem has an unique
strong solution in $C(\RE;D(\Delta^{A,B})\cap
C^1(\RE;D(F^{A,B}))\cap C^2(\RE;\LD)$. Let us denote by $\phi(t)$
such a solution. By the structure of $D(-\Delta^{A,B})$ given in
Lemma \ref{lemma} we know that
$$
\phi(t)=\phi_0(t)+\sum_{1\le j\le n}\zeta^j_\phi(t)\,\G_j\,,
$$
with $\phi_0(t)\in\HDB$ and
\begin{equation}\label{BC}
\sum_{1\le j\le n}A_{ij}\,\phi_0(t,y_j)=\sum_{1\le j\le
n}(B-AG)_{ij}\,\zeta^j_\phi(t)
\end{equation}
for all $t\in\RE$. Let us now define, for any $t\ge 0$,
\begin{equation}\label{decomp}
\phi_f(t):=\phi(t)-\sum_{1\le j\le n}\phi_j(t)\,,
\end{equation}
where $\phi_j(t)$ denotes the spherical wave
$$\phi_j(t):=\theta(t-d_j)\,\zeta_\phi^i(t-d_j)\,\G_j\,.$$
Since
$$
\ddot\phi_j=\Delta\phi_j+\zeta^j_\phi\,\delta_{y_j}\,,
$$
one has
\begin{align*}
\ddot\phi_f=&\Delta\left(\phi_0-\sum_{1\le j\le n}\phi_j\right)
-\sum_{1\le j\le n}\zeta^j_\phi\,\delta_{y_j}
=\Delta\left(\phi_0-\sum_{1\le j\le n}(\phi_j
-\zeta^j_\phi\,\G_j)\right)\\
=&\Delta\left(\phi-\sum_{1\le j\le n}\phi_j\right)=\Delta\phi_f
\end{align*}
and thus $\phi_f$ is the unique solution of the Cauchy problem
(\ref{solf1}), (\ref{solf2}), (\ref{solf3}). Writing the boundary
conditions (\ref{BC}) by using the decompositions of $\phi$ given
by the relation (\ref{decomp}), one obtains
\begin{align*}
&\sum_{1\le j,k\le n}(B_{ij}-A_{ik}G_{kj})\,\zeta^j_\phi(t)\\
=&\sum_{1\le j\le n}A_{ij}\left( \lim_{x\to y_j}
\,\left(\phi(t)-\sum_{1\le k\le n}\zeta_\phi^k(t)\,
\G_k\right)(x)\,\right)\\
=&\sum_{1\le j\le n}A_{ij}\left(\phi_f(t,y_j) +\sum_{1\le k\le
n}\theta(t-d_{jk})\,G_{jk}\,
\zeta^k_\phi{(t-d_{jk})}\right.\\
&\left.-\sum_{1\le k\le n}G_{jk}\zeta^k_\phi(t) +\lim_{x\to y_j}
\,\frac{\zeta^j_\phi(t-|y_j-x|\,)-\zeta^j_\phi(t)}{4\pi|y_j-x|}
\right)\\
=&\sum_{1\le j\le n}A_{ij}\left(\phi_f(t,y_j) +\sum_{1\le k\le
n}\theta(t-d_{jk})
\,G_{jk}\zeta^k_\phi{(t-d_{jk})}\right.\\
&\left.-\sum_{1\le k\le n}\zeta^j_\phi G_{jk}
-\frac{1}{4\pi}\,\dot\zeta^j_\phi(t)\right)\,.
\end{align*}
This shows that $\vec\zeta_\phi$ is a solution of (\ref{retard}),
(\ref{retard1}). Such a Cauchy problem has an unique solution.
Indeed let $\vec\zeta(t)$, with $\vec\zeta(0)=\vec 0$, solve the
system
\begin{equation*}
\frac{1}{4\pi}\,A\dot{\vec\zeta}+B\vec\zeta=0 \,.
\end{equation*}
By (\ref{symm}) and Ker$(A)\cap$Ker$(B)=\{\vec 0\}$, which is a
consequence of (\ref{rang}), one has
$$
|\det(iA+B)|^2=\det((iA+B)(-iA^*+B^*))=\det(AA^*+BB^*)\not=0,
$$
so that $zA+B$, $z\in\C$, is a regular pencil of matrices, i.e.
$p(z):=\det(zA+B)$ is not the zero polynomial. By Theorem 3.2 of
\cite{[Hoo]} this implies $\vec\zeta(t)=0$.
\end{proof}
The above theorem has an analogous version for negative times. In
this case one obtains a representations involving, instead of the
retarded waves $\theta(t-d_j)\,\zeta_\phi^i(t-d_j)\,\G_j$, the
advanced ones $\theta(-t-d_j)\,\zeta_\phi^i(t+d_j)\,\G_j$. Since
there is no substantial difference between these two situations,
in the proof of the following theorem we will consider only the
positive time case, the proof in the negative time case being
essentially the same.\par

The previous theorem shows that compactly supported Cauchy data
always give rise to compactly supported solution. However this
does not necessarily implies a finite speed of propagation. Indeed
we have the following

\begin{theorem} \label{Th3} Let $ - \Delta^{A,B} $ be a point perturbation of the Laplace
operator described by Definition \ref{def1} with the matrices $
A,B $ satisfying (\ref{symm}) and (\ref{rang}). Then the wave
equation
\begin{equation}
\ddot\phi-\Delta^{A,B}\phi=0
\end{equation}
has finite velocity of propagation if and only if $-\Delta^{A,B}$
is defined by local boundary conditions, i.e. if and only if both
matrices $A$ and $B$ can be chosen diagonal simultaneously.
\end{theorem}
\begin{proof} Let $\phi(t)$ be a solution of (\ref{problem}) with
$\diam(S_0)<+\infty$, where
$$
S_0:=\supp \phi(0)\cup\supp\dot\phi(0)\,.
$$
By Theorem \ref{Th2},
$$\phi(t)=\phi_f(t)+\sum_{1\le i\le n}\phi_i(t)\,,
$$
where $\phi_i(t)$ denotes the spherical wave
$$\phi_i(t):=\theta(t-d_i)\,\zeta_\phi^i(t-d_i)\,G_i\,.$$
Since the free wave equation has velocity of propagation equal to
one, we have (here and below $t\ge 0$)
$$
\diam(\supp \phi_f(t))\le 2t+\diam(S_0)
$$
and $$ \diam(\supp \phi_i(t))=\diam(\left\{d_i<t\right\})=2t\,.
$$
Thus, if $Y\subseteq S_0$,
$$
\diam(\supp \phi(t))\le 2t+\diam(S_0)\,.
$$
Let us now consider the case $Y_0:= Y \setminus (S_0\cap Y)
\not=\emptyset$.

 Suppose that $A$ and $B$ can be chosen diagonal. Let $y_i\in Y_0$. By (\ref{retard}), in order the
 wave $\phi_i(t)$ be present, the point
$y_i$ must have been reached at some earlier stage either by the
free wave $\phi_f$ or by a spherical wave $\phi_j$ originated from
a different point $y_j$. Since all these waves travel with the
speed $v=1$, in conclusion one has that
$$
\diam(\supp \phi(t))\le 2t+\diam(S_0)\,.
$$
We have proven sufficiency.

To prove the necessity consider  the operator $ -\Delta^{A,B} $
described by arbitrary matrices $ A,B $ satisfying (\ref{symm})
and (\ref{rang}). Take an arbitrary point $ y_k \in Y $ and
initial data satisfying the following conditions:
\begin{eqnarray}
\label{zeta0}\vec\zeta_{\phi_0}\equiv \vec\zeta_\phi(0) = 0, \\
\dist(y_k,S_0)<\dist(Y\setminus\left\{y_k\right\},S_0).
\end{eqnarray}
Then the solution $ \phi_f $ of (\ref{solf1}), (\ref{solf2}),
(\ref{solf3}) reaches the point $ y_k $ before it reaches any
other point from  $ Y.$  Therefore, for some sufficiently small
$\epsilon>0$ and for any $t$ such that
$\dist(y_k,S_0)<t<\dist(y_k,S_0)+\epsilon$, the system of
equations (\ref{retard}) takes the form
\begin{equation} \label{retard2}
\sum_{1\le j\le n}\frac{1}{4\pi}\,A_{ij}\dot\zeta_\phi^j+
B_{ij}\zeta_\phi^{j} =A_{ik}\phi_f(y_k)\,,\qquad i=1,....,n\,.
\end{equation}
The evolution has finite speed of propagation only if all $
\zeta_\phi^i, i \neq k $ are zero, otherwise
$$
\diam(\supp\phi(t))>
|y_i-y_k|\,,\quad\dist(y_k,S_0)<t<\dist(y_k,S_0)+\epsilon
$$
which violates (\ref{lsp}) since both $\diam(S_0)$ and
$\dist(y_k,S_0)$ can be made arbitrarily.\par If all
$\zeta_\phi^i$ are zero for $i\not=k$ then (\ref{retard2}) becomes
$$
\frac{1}{4\pi}\,A_{ik}\dot\zeta_\phi^k+ B_{ik}\zeta_\phi^{k}
=A_{ik}\phi_f(y_k)\,,\qquad i=1,....,n\,,
$$
so that the $k$-th columns of $A$ and $B$ are linearly dependent.
Since the point $y_k$ is chosen arbitrarily from $Y$, finite speed
of propagation implies that the linear spaces $ V_k $ spanned by
the $k$-th columns of $A$ and $B$ are one-dimensional for all $k$.
Then condition (\ref{rang}) implies that any vector $ v =
\sum_{k=1}^n v_k, \, v_k \in V_k $ is equal to zero only if all $
v_k = 0.$ Therefore the corresponding Lagrangian plane is a tensor
product of Lagrangian planes for boundary values associated with
each singular point $ y_k$, i.e. that the boundary conditions are
local.

\end{proof}

\end{document}